\documentclass[12pt]{article}
\usepackage{times,color}
\usepackage{amsmath}
\usepackage{amssymb}
\usepackage{xcolor} 
\usepackage{subfigure,epsfig,graphicx,epstopdf}
\usepackage[colorlinks]{hyperref}
\hypersetup{linkcolor = {black}, citecolor = {black}, urlcolor = {black}}
\usepackage[
    backend=biber,
    style=nature,
    sorting=none
]{biblatex}

\AtEveryBibitem{\clearfield{language}}
\AtEveryBibitem{\clearfield{issn}}
\AtEveryCitekey{\clearfield{issn}}
\AtEveryBibitem{\clearfield{url}}
\AtEveryBibitem{\clearfield{month}}
\AtEveryBibitem{\clearfield{urldate}}
\AtEveryBibitem{\clearfield{date}}

\usepackage{upgreek}

\usepackage{caption}
\DeclareCaptionLabelFormat{bflabel}{\textbf{#1~#2}} % 自定义加粗格式
\newenvironment{sciabstract}{%
\begin{quote} \bf}
{\end{quote}}

\newcounter{lastnote}

\begin{document}

\title{Spatiotemporally Interleaved Homodyne Photonic Tensor Core}
\author{Yun-Long Nie$^{1,3}$, Hang Song$^{1,3}$, De-Hui Huang$^{1,3}$, Yi Xie$^{1,3}$, Yu-Xuan Fu$^{1,3}$, \\Jian-Peng Dou$^{1,3}$, Xiao-Yun Xu$^{1,3}$, Hao Tang$^{1,3,4,\ast}$, Xian-Min Jin$^{1,2,3,4,\ast}$\\
\\
\normalsize{$^1$Center for Integrated Quantum Information Technologies (IQIT), School of Physics}\\ \normalsize{and Astronomy and State Key Laboratory of Photonics and Communications,}\\
\normalsize{Shanghai Jiao Tong University, Shanghai 200240, China}\\
\normalsize{$^2$TuringQ Co., Ltd., Shanghai 200240, China}\\
\normalsize{$^3$Hefei National Laboratory, Hefei 230088, China}\\
\normalsize{$^4$Chip Hub for Integrated Photonics Xplore (CHIPX),}\\ 
\normalsize{Shanghai Jiao Tong University, Wuxi 214000, China}\\
\normalsize{$^\ast$E-mail: htang2015@sjtu.edu.cn, xianmin.jin@sjtu.edu.cn}}
\date{}%\today
%\begin{document}
% Double-space the manuscript.
\baselineskip24pt

\maketitle

\begin{sciabstract}
Photonic computing provides ultrahigh bandwidth, low latency and intrinsic parallelism, making it a promising route beyond the scaling limits of electronic computing. However, existing on-chip photonic computing systems remain constrained by persistent trade-offs among high-speed modulation, energy efficiency and large-scale integration, limiting their system-level advantages. Here we present a spatiotemporally interleaved homodyne photonic tensor core implemented on a thin-film lithium niobate (TFLN) platform. By integrating a homodyne photonic matrix with a bus-readout time-integrating array, this architecture scales down the high-speed digital-to-analog and electro-optic interconversion hardware overhead required for photonic matrix operations from O(\textit{n}$^2$) to O(\textit{n}), thereby unlocking system-level scalability. Moreover, the architecture employs orthogonal horizontal and vertical crossbars to route data and weight signals independently, eliminating the intrinsic beam combining loss while enabling ultrahigh-speed synchronous updates of both data and weights. Collectively, these features provide a scalable and hardware-efficient foundation for high-bandwidth photonic processors targeting general-purpose artificial intelligence (AI) computing.
\\
\end{sciabstract}

\section*{Introduction}
\noindent 

The rapid expansion in the scale and computational complexity of artificial intelligence (AI) models \cite{Mehonic2022,Xu2024} is imposing increasingly stringent system-level demands on computing hardware. As model parameters, context lengths and multimodal input dimensions to grow, hardware platforms must deliver more than higher peak throughput; they also require to balance communication bandwidth and energy efficiency across data loading, weight updating, on-chip interconnection and result readout. Photonic computing \cite{Yan2025,Zhang2025,Cheng2024,Xue2024,Chen2025,Fu2026} exploits the high-speed modulation, low latency and interference-enabled parallelism of light to execute linear operations directly, offering a promising route towards high-bandwidth AI accelerators. Especially, in on-chip photonic matrix processors \cite{Hu2025,Zhang2025_Thin,Dong2024,Li2024}, input data and weights can be encoded into optical degrees of freedom, including intensity, phase \cite{Onodera2025}, wavelength \cite{Luan2026,Zhang2025_Unlocking,feldmann_parallel_2021}, space \cite{Cheng2024_Photonic} and time \cite{Dong2023}, while matrix operations and result readout are implemented through optical-field superposition and photodetection. This approach provides a powerful hardware foundation for large-scale parallel computing \cite{sun2026fully}.

However, as photonic matrix computing progresses from small-scale demonstrations towards large-scale AI workloads, the dominant bottleneck is shifting from the modulation speed of individual photonic devices to the scalability of system-level resources \cite{Ahmed2025,Hua2025}. First, when each weight or multiplication node in the matrix requires independent high-speed modulation, the number of modulators, digital-to-analog converters (DACs) and driving interconnects scales as O(\textit{n}$^2$) with the matrix size \textit{n}, substantially increasing write-interface complexity and power consumption \cite{Lam2026,Wang2025,Ning2025,Ou2025}. Second, conventional crossbar architectures rely on optical signal combining for accumulation, leading to an input-size-dependent beam combining loss that substantially increases the optical power budget as the matrix dimension increases \cite{BrckerhoffPlckelmann2024}. Finally, even when optical multiplication is performed with high parallelism, the resulting O(\textit{n}$^2$) outputs must still be detected by high-speed photodetectors and sampled by analog-to-digital converters (ADCs) \cite{feldmann_parallel_2021,Ou2025}. The number of ADCs, their sampling rates and the associated electrical interconnects therefore impose additional scaling bottlenecks at the readout interface. Existing on-chip photonic matrix processors thus still lack a system-level architecture that can preserve O(\textit{n}$^2$) parallel multiply-accumulate operations while reducing both write and readout high-speed electrical interfaces to O(\textit{n}), and simultaneously avoids the power penalty in optical beam combining.

Here, we propose a spatiotemporally interleaved homodyne photonic tensor core (HPTC) implemented on a thin-film lithium niobate (TFLN) \cite{Song2025,Han2026,Zhu2021} platform (Fig. 1). The architecture introduces orthogonal horizontal and vertical modulation \cite{Bernstein2021} at the write interface, enabling computing units within the same column of the photonic matrix to share one set of weights. This scheme reduces the number of write-side electro-optic modulators and DACs from O(\textit{n}$^2$) to O(\textit{n}) without compromising computational generality, thereby substantially relaxing the bandwidth requirement for data writing. At the compute interface, coherent homodyne detectors \cite{RahimiKari2024,chen_deep_2023,li2025hybrid,zhu2025enlighten,11004625,Tian2025} combined with time integrators serve as photonic matrix elements that locally perform multiply-accumulate operations. This approach eliminates the intrinsic beam combining loss of conventional crossbar architectures \cite{Jiang2025}, reducing the system optical power budget from O(\textit{n}$^3$) to O(\textit{n}$^2$). At the readout interface, time-domain accumulation significantly lowers the sampling rate required for matrix operations, thus alleviating data-readout bandwidth demands, while a bus-based readout circuit further scales down the number of ADC channels from O(\textit{n}$^2$) to O(\textit{n}). Combined with the high-speed electro-optic modulation enabled by TFLN, this architecture effectively addresses the long-standing trade-off among high bandwidth, high energy efficiency and large-scale integration. Moreover, the synergy between orthogonal modulation and homodyne detection enables high-speed dynamic updates of both data and weights, allowing the architecture to efficiently map the massive parameters and high-dimensional matrices required by large-scale AI models.
\\
\\
\noindent \textbf{Chip architecture}

The HPTC consists of a homodyne-crossbar photonic matrix and a bus-readout time-integrating array. The homodyne-crossbar photonic matrix performs data encoding and multiplication (Fig. 1a). The operation begins by equally splitting the input laser into the data and weight modules. Two columns of modulators, arranged as an \textit{n} × 2 array, encode the input data \textbf{X} and weights \textbf{W} into their corresponding spatiotemporal domains (\textit{n} × \textit{t}), with each modulator carrying an independent optical signal stream. The resulting \textit{n} data streams and \textit{n} weight streams are then routed into an \textit{n} × \textit{n} crossbar photonic matrix through horizontal row lines and vertical column lines, respectively. Within the matrix, the optical signals from the \textit{n} rows and \textit{n} columns are uniformly distributed to \textit{n} × \textit{n} matrix elements. At each element, the data and weight streams are multiplied through coherent homodyne detection. Specifically, they are directed into a multimode interferometer (MMI) for interference, producing two optical output signals. The two outputs are then converted into photocurrents by photodetectors, whose difference yields the multiplication result, $i_u - i_d \propto xw$ (Fig. 1b and Supplementary Information 1).

The bus-readout time-integrating array performs data accumulation and readout (Fig. 1c and Methods). Time-domain accumulation markedly decreases the number of sampling events. For example, for a matrix operation with a length of 4096, the required readout bandwidth can be lowered by a factor of 4096. Therefore, a bus-based readout scheme can be introduced, in which ADCs are reused over time. The circuit implementation of a single matrix-element time integrator is shown in Fig. 1d. The multiplication result, generated as an electrical signal, is first accumulated on the integration capacitor C$_1$, then transferred to the storage capacitor C$_2$, and finally read out row by row. Switch S$_1$ defines the accumulation window for the input signal, S$_2$ resets the accumulated signal, S$_3$ transfers the accumulated signal, S$_4$ resets the stored signal, and S$_5$ controls row-wise readout from storage node. By decoupling the integration capacitor from the storage capacitor, the transfer switch S$_3$ allows accumulation in the current stage and readout of the previous stage to proceed independently, thereby enabling continuous operation. Switches S$_1$-S$_4$ share global control signals across the array, whereas S$_5$ is controlled row by row. Consequently, accumulation and transfer are globally synchronized, while readout is synchronized within each row. This operating sequence collectively implements matrix-matrix multiplication (Fig. 1e):
\[\textbf{Y}(n\times n)=\textbf{W}(n\times t)\textbf{X}(t\times n).\]

The HPTC provides an architectural route to mitigate the trade-off among high-speed modulation, energy efficiency and large-scale integration. A conventional photonic matrix contains \textit{n}$^2$ independent weight cells and performs the multiplication between a weight matrix \textbf{W} and a data vector \textbf{x}, namely \textbf{y = Wx}, within a single clock cycle. In contrast, the HPTC employs orthogonal horizontal and vertical modulation, allowing the computing units in each column to share the same set of weights. This enables the architecture to compute the outer product between a weight vector \textbf{w} and a data vector \textbf{x}, $\textbf{y}=\textbf{w}\otimes\textbf{x}$, within one clock cycle, thereby reducing the number of electro-optic modulators and ADCs from \textit{n}$^2$+\textit{n} to 2\textit{n}. By combining vector outer products with time-domain accumulation, the HPTC implements full matrix-matrix multiplication without sacrificing computational generality. Unlike mainstream spatial-domain multiply-accumulate schemes, this architecture employs homodyne detectors and time integrators as the fundamental computing units, thereby performing multiply-accumulate operations in the time domain. This design avoids multi-channel beam combining, eliminates the input-size-dependent intrinsic loss, and scales the input optical power budget from \textit{n}$^3$ to \textit{n}$^2$. Meanwhile, by combining time-domain accumulation with bus-based readout, the architecture relaxes the readout requirement from \textit{n}$^2$ high-speed ADCs to only 2\textit{n} low-speed ADCs. As a result, the architecture relaxes, across data writing, computation and readout, the dependence of large-scale optoelectronic computing on dense arrays of high-speed digital-analog interconversion and electro-optic interconversion devices, as well as on a large optical power budget. For a fixed computational scale, it therefore reduces the required data bandwidth, high-speed interconnect complexity and energy cost with power-law scaling.
\\
\\
\noindent \textbf{Fabrication of the chip}

The HPTC chip was fabricated in-house on a 6-inch lithium-niobate-on-insulator (LNOI) wafer using a wafer-scale integration process, as shown in Fig. 2a,b. As a proof of concept demonstration, the core functional blocks of the HPTC, including electro-optic modulators, a crossbar photonic matrix and homodyne detectors, were integrated on a single chip. The prototype implements a 4 × 4 array layout with an overall footprint of 11.7 × 4.5 mm$^2$. The input optical carrier is first coupled into the chip via a dual-layer spot-size converter with a mode-field diameter of 6.4 \textmu m, and is then uniformly distributed to the data- and weight-encoding units through a three-stage binary-tree cascade of MMIs. To enable compact, high-speed on-chip encoding, three-section folded travelling-wave electrode modulators are employed for both data and weight modulation units. At the folded bends, the waveguides in the upper and lower arms are crossed to maintain a consistent modulation direction across all sections. The signal electrode is further extended to match the electrical propagation delay with the optical delay. This folded configuration compresses a 7.0-mm-long modulation electrode into a device length of only 2.8 mm, thereby improving the integration density. The folded modulators encode the electrical data and weight signals onto the optical carriers, which are subsequently routed into the crossbar photonic matrix through horizontal row lines and vertical column lines to perform parallel multiplication.

In the homodyne crossbar, the optical signals are evenly distributed to each homodyne-detection matrix element through a set of directional couplers. The splitting ratio of the \textit{i}-th row or column coupler is designed as 1/(5-\textit{i}), with \textit{i} denoting the row or column index. Because the multiplication results are accumulated by time integration, strictly identical optical power at all matrix elements is not required. This property relaxes the requirement on splitting-ratio accuracy and improves the tolerance of the homodyne crossbar to fabrication-induced variations. Each homodyne detector consists of a thermo-optic phase shifter, an MMI, two grating couplers and two photodetectors. The thermo-optic phase shifter adjusts the relative phase between the two input signals, allowing the MMI to operate at the optimal constructive/destructive interference condition. The MMI then coherently combines the two inputs and generates two outputs corresponding to constructive and destructive interference, respectively. These outputs are then vertically coupled out of the chip surface through the grating couplers and collected by flip-chip-bonded InGaAs photodetectors, where they are converted into electrical signals for balanced detection. The InGaAs photodetectors used in this work have a responsivity of 0.95 A/W. Finally, the HPTC chip is electrically interfaced by wire bonding and optically coupled via a fibre array (Fig. 2c).
\\
\\
\noindent \textbf{Characterization of the chip}

The electro-optic bandwidth of the folded modulator was characterized using a 67 GHz vector network analyzer (R\&S ZNA67EXT). The normalized electro-optic response shows that the folded modulator achieves a -3 dB bandwidth of 37.8 GHz (Fig. 2d), confirming its capability for high-speed modulation. We then evaluated the analog computing performance of the chip. Through coherent homodyne detection, the two output optical signals are converted into electrical signals and differentially read out, yielding a differential current $i_u - i_d \propto xw$, where $x$ and $w$ denote the amplitudes of the data and weight signals encoded onto the optical carriers, respectively. In contrast to the intensity encoding commonly used in conventional photonic processors, this amplitude-encoding scheme naturally supports both positive and negative values, thereby enabling real-valued multiplication. To quantify computing accuracy, we tested a single matrix element by sweeping the data and weight amplitudes over \textbf{x},\textbf{w}=[-1:0.1:1]\textsuperscript{T}, corresponding to a real-valued vector outer product $\textbf{x}\otimes\textbf{w}$ and forming a 21×21 multiplication-table benchmark. At a modulator rate of 10 Mbaud, the experimental results are compared with the theoretical values in Fig. 2e. The scatter plot shows that the measured data points cluster near the diagonal, and residual analysis yields a standard deviation of 0.0621, quantifying the computing error of this matrix element (Fig. 2f).
\\

\section*{Conclusion}
\noindent 

In this work, we demonstrate a TFLN spatiotemporally interleaved homodyne photonic tensor core. By combining a homodyne photonic matrix with a bus-readout time-integrating array, the architecture reduces the key device resources required at the write, compute and readout interfaces from O(\textit{n}$^2$) to O(\textit{n}) without compromising the computational scale or generality of photonic matrix operations, leading to a substantial improvement in system-level hardware efficiency, and thereby unlocking the scalability of photonic computing for high-bandwidth and energy-efficient applications. The fabricated HPTC chip achieves an electro-optic bandwidth of 37.8 GHz, and benchmark measurements show a computation error of 0.0621. These results establish a promising route towards high-performance photonic computing systems for general-purpose AI.

\textit{Notes added}. We have been made aware of a related but not overlapping work \cite{zhou2026homodyne} recently.
\\
\section*{Methods}
\noindent \textbf{The operating sequence of the bus-readout time-integrating array.} 

During initialization, (1) switch S$_2$ is closed and then opened to reset the integration capacitor C$_1$; (2) switch S$_4$ is closed and then opened to reset the storage capacitor C$_2$. During accumulation, (3) switch S$_1$ is close to initiate accumulation, allowing the charge signal generated by the differential photodetectors to accumulate on C$_1$; (4) S$_1$ is opened to terminate the integration window; (5) S$_3$ is closed and then opened to transfer the accumulated charge from C$_1$ to C$_2$; and (6) S$_2$ is closed and then opened to reset C$_1$. The accumulation process is then repeated by cycling through steps (3)-(6). During readout, (7) after step (5), the transferred charge is stored on C$_2$, which is isolated from the bus by a voltage follower to prevent disturbance of the stored signal during direct readout; (8) the row-selection switch S$_5$ is sequentially closed and opened row by row, delivering the stored signal to the column output bus and enabling row-wise readout of the accumulated signal array; and (9) S$_4$ is closed and then opened to reset C$_2$. The readout sequence is then repeated by cycling through steps (7)-(9). In this circuit, each switch can be implemented using a single transistor.
\\
\\
\printbibliography{}

\baselineskip21pt
\clearpage

%===========================fig1==========================
\begin{figure*}
\centering
\includegraphics[width=1 \columnwidth]{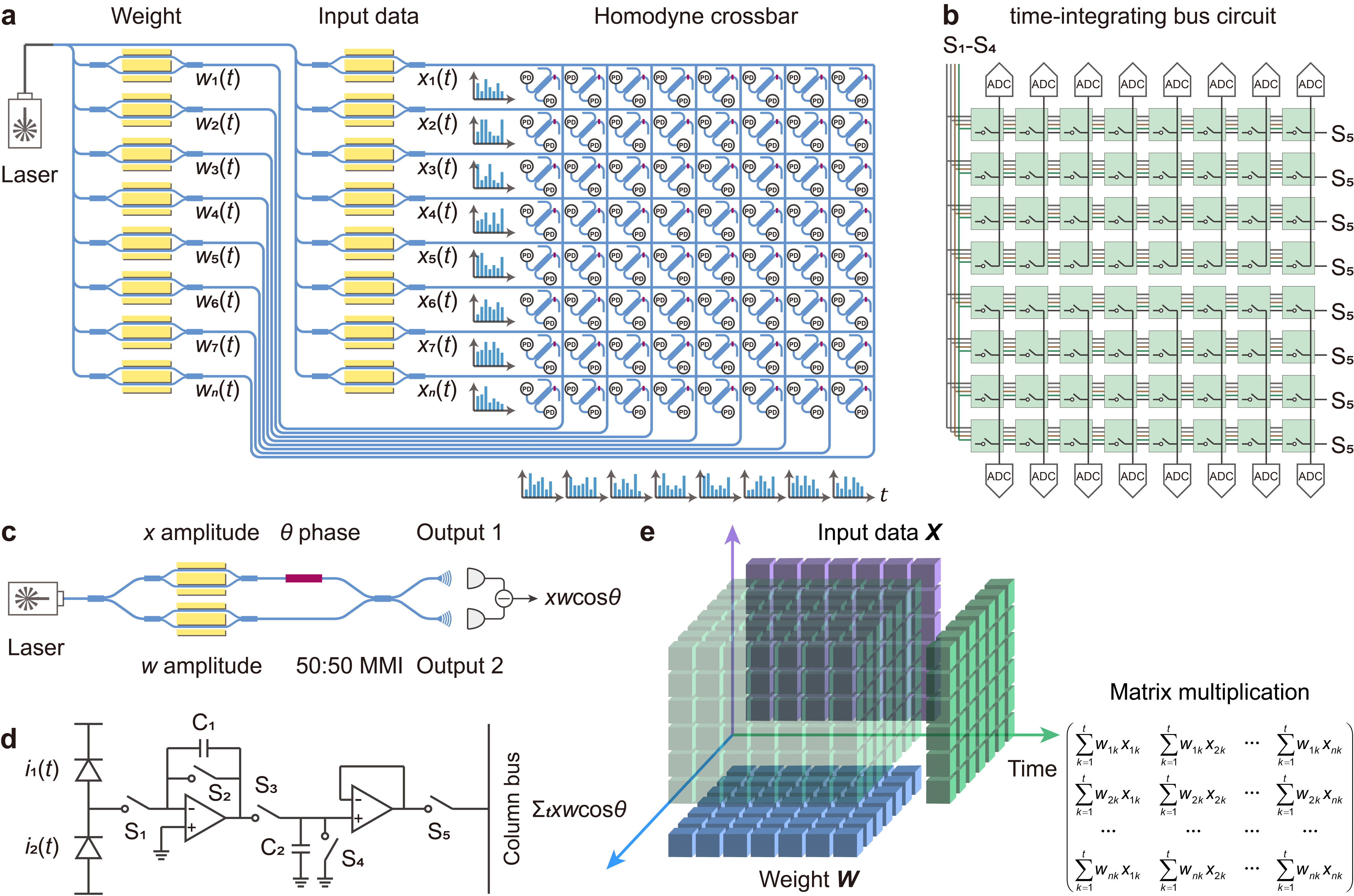}
\caption{\textbf{Overview and working principle of the spatiotemporally interleaved homodyne photonic tensor core}. \textbf{a}, Homodyne-crossbar photonic matrix for data encoding and multiplication. \textbf{b}, Homodyne balanced detection, in which the data and weight optical signals interfere to generate a differential photocurrent proportional to their product, $i_u - i_d \propto xw$. \textbf{c}, Bus-readout time-integrating array for data accumulation and readout. \textbf{d}, Time-integrator circuit. \textbf{e}, Matrix-matrix multiplication between the data matrix \textbf{X} and the weight matrix \textbf{W}.}
\label{fig1}
\end{figure*}
%=====================================================

%============================fig2===========================
\begin{figure*}
\centering
\includegraphics[width=1 \columnwidth]{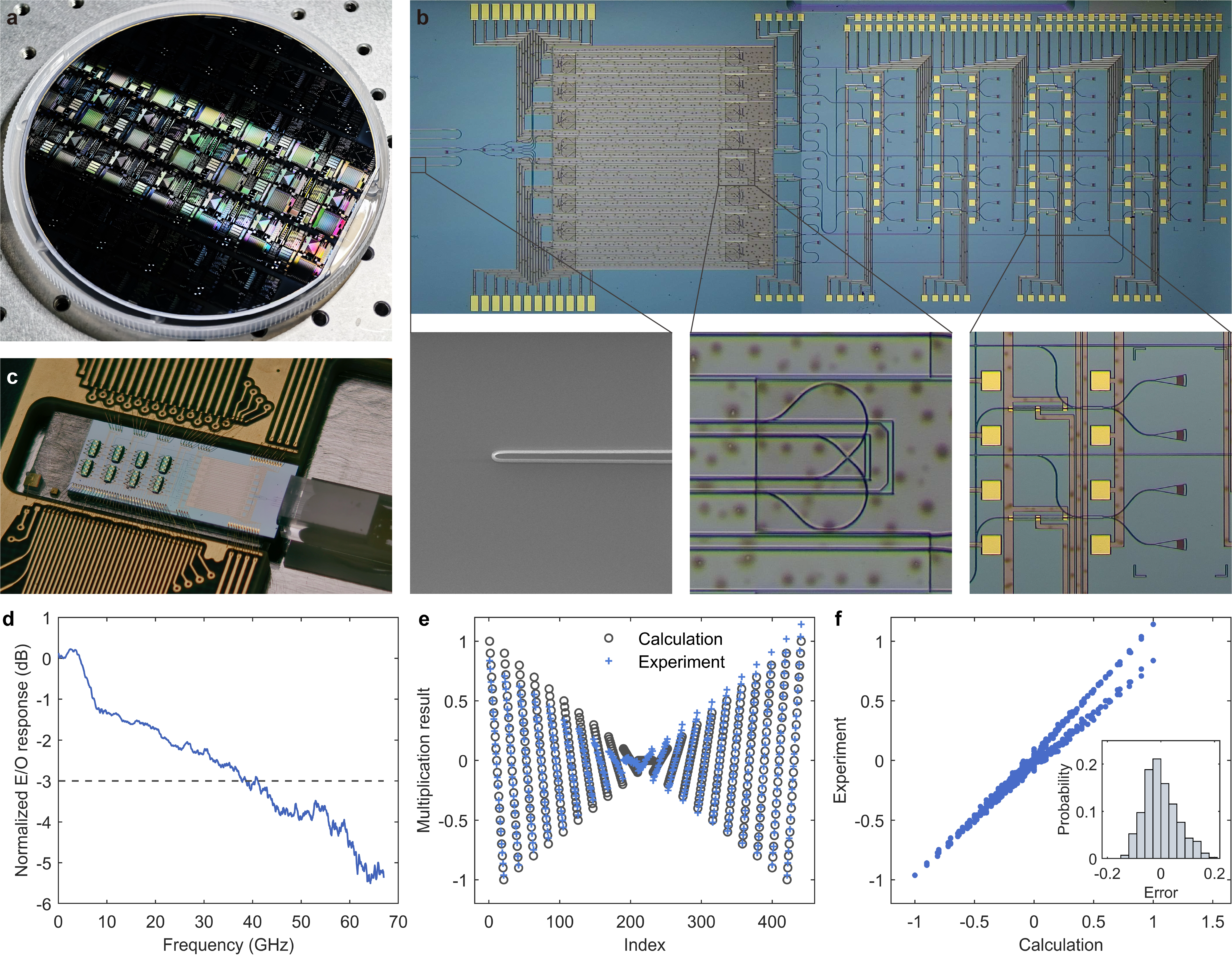}
\caption{\textbf{Fabrication and characterization of the HPTC chip}. \textbf{a}, Wafer-scale fabrication of TFLN HPTC chips on a 6-inch wafer, containing 24 complete shots, each with an area of 22 × 22 mm$^2$. \textbf{b}, Optical micrograph of the chip. Insets show the 6.5-\textmu m mode-size converter, the folded section of the modulator and the homodyne detector. \textbf{c}, Packaged 4 × 4 HPTC chip with fibre-array coupling and wire-bonded electrical interfaces. \textbf{d}, Measured electro-optic S21 response of the modulator. \textbf{e}, Benchmark results for real-valued multiplication tables. \textbf{f}, Scatter plot of the experimental results corresponding to the benchmark in \textbf{e}.
}
\label{fig2}
\end{figure*}
%=====================================================

\end{document}